\documentclass[10pt,conference ]{IEEEtran}
\usepackage{amsfonts}
\usepackage{times}
\usepackage{graphicx}
\usepackage{latexsym}
\usepackage{dsfont}
\usepackage{amssymb}
\usepackage{amsmath}
\usepackage{cite}
\usepackage{verbatim}
\usepackage{subfigure}

\def\bb0{{\mathbb{0}}}

\def\bb{{\mathbf{b}}}

\def\b0{{\mathbf{0}}}

\def\bS{{\mathbf{S}}}

\def\bX{{\mathbf{X}}}

\def\sf0{{\mathsf{0}}}

\usepackage{multirow}
\usepackage{multicol}
\usepackage{booktabs}
\usepackage{epstopdf}
\usepackage{enumerate}
\usepackage{algorithm}
\usepackage{amsmath}
\usepackage{float}
\usepackage{color}
\usepackage{makeidx}
\usepackage{bm}
\usepackage{cleveref}
\usepackage{url}
\usepackage{steinmetz}
\usepackage{varwidth}
\usepackage{soul}
\graphicspath{{figs/}}
\usepackage{tabularx}

\newcommand{\comm}[1]{}

\begin{document}

\title{Environment Semantic Aided Communication: \\ A Real World Demonstration for Beam Prediction  }
\author{Shoaib Imran, Gouranga Charan, and Ahmed Alkhateeb\\ \textit{School of Electrical, Computer, and Energy Engineering, Arizona State University} \\ Emails: \{s.imran, gcharan, alkhateeb\}@asu.edu }

\maketitle

\begin{abstract}
Millimeter-wave (mmWave) and terahertz (THz) communication systems adopt large antenna arrays to ensure adequate receive signal power. However, adjusting the narrow beams of these antenna arrays typically incurs high beam training overhead that scales with the number of antennas. Recently proposed vision-aided beam prediction solutions, which utilize \textit{raw RGB images} captured at the basestation to predict the optimal beams, have shown initial promising results. However, they still have a considerable computational complexity, limiting their adoption in the real world. To address these challenges, this paper focuses on developing and comparing various approaches that extract  lightweight semantic information from the visual data. The results show that the proposed solutions can significantly decrease the computational requirements while achieving similar beam prediction accuracy compared to the previously proposed vision-aided solutions.

\end{abstract}


\section{Introduction} \label{sec:Intro}
Future communication systems are shifting to higher frequency bands like mmWave in 5G and potentially sub-terahertz in 6G and beyond. These bands offer high bandwidth, enabling the communication systems to support the increasing data rate demands of new applications like autonomous driving, 8K video streaming, and mixed reality \cite{Rappaport2019}. However, these systems require large antenna arrays and use narrow directive beams for both the transmitter and receiver to ensure sufficient receive signal power. Selecting the optimal beams for these arrays results in a large beam training overhead. This high beam training overhead makes it challenging to support highly mobile and low-latency applications, making it a key challenge in deploying these systems. Therefore, there is a need to find new ways to reduce this training overhead and enable highly-mobile mmWave/THz communication systems.

Several solutions to reduce the beam training and channel estimation overhead in mmWave/THz communication systems have been proposed over the years \cite{Alkhateeb2014,Jayaprakasam2017,HeathJr2016}. These solutions have included constructing adaptive beam codebooks \cite{Zhang2021b}, designing beam tracking techniques \cite{Jayaprakasam2017}, and leveraging the channel sparsity and compressive sensing tools \cite{Alkhateeb2014,HeathJr2016}. While these classical approaches were able to achieve some improvement, they typically only save one order of magnitude in the training overhead, which is not enough for systems with large antenna arrays, particularly for serving highly-mobile and low-latency wireless communication applications.

The challenges faced by classical approaches have led to the development of machine learning-based (ML) solutions to address the beam training and channel estimation overhead in mmWave/THz communication systems \cite{Morais22, charan2021c,Jiang_LiDAR,demirhan2021beam}. For instance, sensory data such as user position and orientation \cite{Morais22}, RGB images \cite{charan2021c}, LiDAR point clouds \cite{Jiang_LiDAR}, and radar measurements \cite{demirhan2021beam} can be utilized to reduce the overall training overhead. Recent work on vision-aided beam prediction has shown initial promising results \cite{Charan_TxID, charan2021c}. The current solutions use raw RGB images and CNN-based architectures to predict the optimal beams. Utilizing the raw RGB images for downstream wireless communication tasks results in higher storage requirements and increased computational cost, limiting their applicability in the real world.

Instead of directly using the raw RGB images, one promising solution can be to extract environment semantics from the images and then utilize that information to predict the optimal beam indices. Environment semantics in the form of image masks, bounding boxes, etc., present several benefits over raw RGB images. Specifically, these semantics only consist of relevant information such as object class and location, and it helps in reducing the storage cost. Further, the lower complexity of the semantics helps to reduce the computation requirements in the downstream task. An important question is whether environment semantics aided beam prediction solutions can achieve similar performance as that in \cite{Charan_TxID, charan2021c}. In this paper, we attempt to answer this important question. The main contributions can be summarized as follows:
\begin{itemize}
	\item {Formulating the environment semantic aided beam prediction problem for mmWave and THz communication systems considering practical sensing/visual and communication models.}
	\item{Developing a deep learning based solution for mmWave/THz beam prediction that utilizes different environment semantics such as image masks and bounding boxes of the detected objects}
	\item{Providing the first real-world evaluation of environment semantic-aided beam prediction based on our large-scale dataset, DeepSense 6G \cite{DeepSense}, that consists of co-existing multi-modal sensing and wireless communication data.}
	\item{Comparing the performance of various semantic-based approaches in terms of beam prediction accuracy and computational complexity, and making important conclusions on which semantic could be more useful in practice. }
\end{itemize}
Based on the adopted real-world datasets, the developed solution can significantly reduce the computational complexity while achieving similar beam prediction accuracy as compared to the prior vision-based solutions. 


\section{Environment Semantic Aided Beam Prediction: \\ System Model and Problem Formulation }\label{sec:sys_ch_mod}
This work considers a communication system where a mmWave basestation is serving a mobile user (vehicle) in a real wireless communication environment. In this section, we first present the adopted wireless communication system model. Next, we formulate the environment semantics-aided beam prediction problem.

\subsection{System Model} \label{sec:sys_model}
This paper adopts the system model, where a basestation, equipped with an $M$-element uniform linear array (ULA) and an RGB camera, is serving a mobile user. The user carries a single-antenna transmitter and is equipped with a GPS receiver capable of collecting real-time position information. The adopted communication system employs OFDM transmission with K subcarriers and a cyclic prefix of length D. To serve the mobile user, the basestation is assumed to employ a pre-defined beamforming codebook $\boldsymbol{\mathcal F}=\{\mathbf f_q\}_{q=1}^{Q}$, where $\mathbf{f}_q \in \mathbb C^{M\times 1}$ and $Q$ is the total number of beamforming vectors. Let $\mathbf h_{k}[t] \in \mathbb C^{M\times 1}$ denote the channel between the basestation and the mobile user at the $k$th subcarrier and time $t$, then the downlink received signal at the user can be written as 
\begin{equation}\label{eq:sys_mod}
	y_{k}[t] = \mathbf h_{k}^T[t] \mathbf f_q[t]x + v_k[t],
\end{equation}
where $\mathbf f \in \boldsymbol{\mathcal F}$ is the optimal beamforming vector at time $t$ and $v_k[t]$ is a noise sample drawn from a complex Gaussian distribution $\mathcal N_\mathbb C(0,\sigma^2)$. The transmitted complex symbol $x\in \mathbb C$  satisfies the power constraint $\mathbb E\left[ |x|^2 \right] = P$, where $P$ is the average symbol power. The beamforming vector $\mathbf f^{\star}[t] \in \boldsymbol{\mathcal F}$ at each time step $t$ is selected to maximize the average receive SNR and is defined as 
\begin{equation}\label{eq:beam_training}
	\mathbf f^{\star}[t] = \underset{\mathbf f_q[t]\in \boldsymbol{\mathcal F}}{\text{argmax}} \frac{1}{K}\sum_{k=1}^{K} \mathsf{SNR}|\mathbf h_{k}^T[t] \mathbf f_q[t] |^2,
\end{equation}
where $\mathsf{SNR}$ is the transmit signal-to-noise ratio, SNR = $\frac{P}{\sigma^2}$.

\subsection{Problem Formulation} \label{sec:prob_form}

Given the system model in Section~\ref{sec:sys_model}, at any given time instant $t$, the task of beam prediction can be defined as selecting the optimal beamforming vector $\mathbf f^{\star}[t] \in \boldsymbol{\mathcal F}$ that maximizes the average receive power. As presented in \eqref{eq:beam_training}, computing the optimal beam indices require explicit channel knowledge, which is, in general, hard to acquire. One other way is to perform exhaustive search over the pre-defined beam codebook. However, the mmWave/THz  communication systems need to deploy large antenna arrays and use narrow directed beams to guarantee sufficient receive SNR. Selecting the optimal beams for these systems with large antenna arrays through exhaustive search is typically associated with large training overhead; making it challenging for these systems to support high mobility wireless communication applications. One promising way to reduce the large beam training overhead is to develop machine learning-based solution that leverages prior observation and additional side information for fast mmWave/THz beam prediction. Recent work on sensing-aided beam prediction have achieved initial promising results by utilizing sensory data such as GPS positions \cite{Morais22}, RGB images \cite{charan2021c}, LiDAR point clouds \cite{Jiang_LiDAR} and radar observations \cite{demirhan2021beam}. In this paper, we propose to utilize additional sensory data (RGB images captured by the basestation) to predict the optimal index for the transmitter in the scene. However, instead of using the raw images captured by the basestation, we propose to extract environment semantics (object masks, bounding boxes, etc.) from the RGB images. These extracted semantics can then be used to predict the optimal beam indices.  

In this work, we target predicting the optimal beam indices based on the availability of RGB images captured by a camera installed at the basestation. Formally, we define $\bX[t] \in \mathbb{R}^{W \times H \times C}$ as the corresponding RGB image, captured by a camera installed in the basestation at time $t$, where $W$, $H$, and $C$ are the width, height, and the number of color channels of the image. Let, $\bS[t]$ represent the environment semantics extracted from the visual data captured by the mmWave/THz basestation. The objective of the beam prediction task is to find a mapping function $f_{\Theta}$ that utilizes the available semantics, $\bS[t]$ to predict  the (estimate) optimal beam index $ \hat{\mathbf f}[t] \in \boldsymbol{\mathcal F}$ with high fidelity. The mapping function can be formally expressed as
\begin{equation}
	f_{\Theta}: \bS[t] \rightarrow  \hat{\mathbf f}[t].
\end{equation}
In this work, we develop a machine learning model to learn this prediction function $f_{\Theta}$. Let $ \mathcal D = \left\lbrace \left (\bS_u, \mathbf f^{\star}_u \right) \right\rbrace_{u=1}^U $ represent the available dataset consisting of image-beam pairs is collected from the real wireless environment, where $U$ is the total number of samples in the dataset. Then, the goal is to maximize the number of correct predictions over all the samples in the dataset $ \mathcal D$. This can be formally written as 
\begin{equation}\label{eq:prob_form_1}
	f^{\star}_{\Theta^{\star}} = \underset{f_{\Theta}}{\text{argmax}}\\ \prod_{u=1}^U \mathbb P\left( \hat{\mathbf f}_u = \mathbf f^{\star}_u | \bS_u \right).
\end{equation}
The prediction function is parameterized by a set of model parameters $\Theta$ and is learned from the labeled data samples in the dataset $\mathcal{D}$. The objective is to find the best parameters $\Theta^{\star}$ that maximize the product of the probabilities of correct predictions. Next, we present our proposed machine learning-based solution for environment semantics-aided mmWave/THz beam prediction.


\begin{figure*}[!t]
	\centering
	\includegraphics[width=0.9\linewidth]{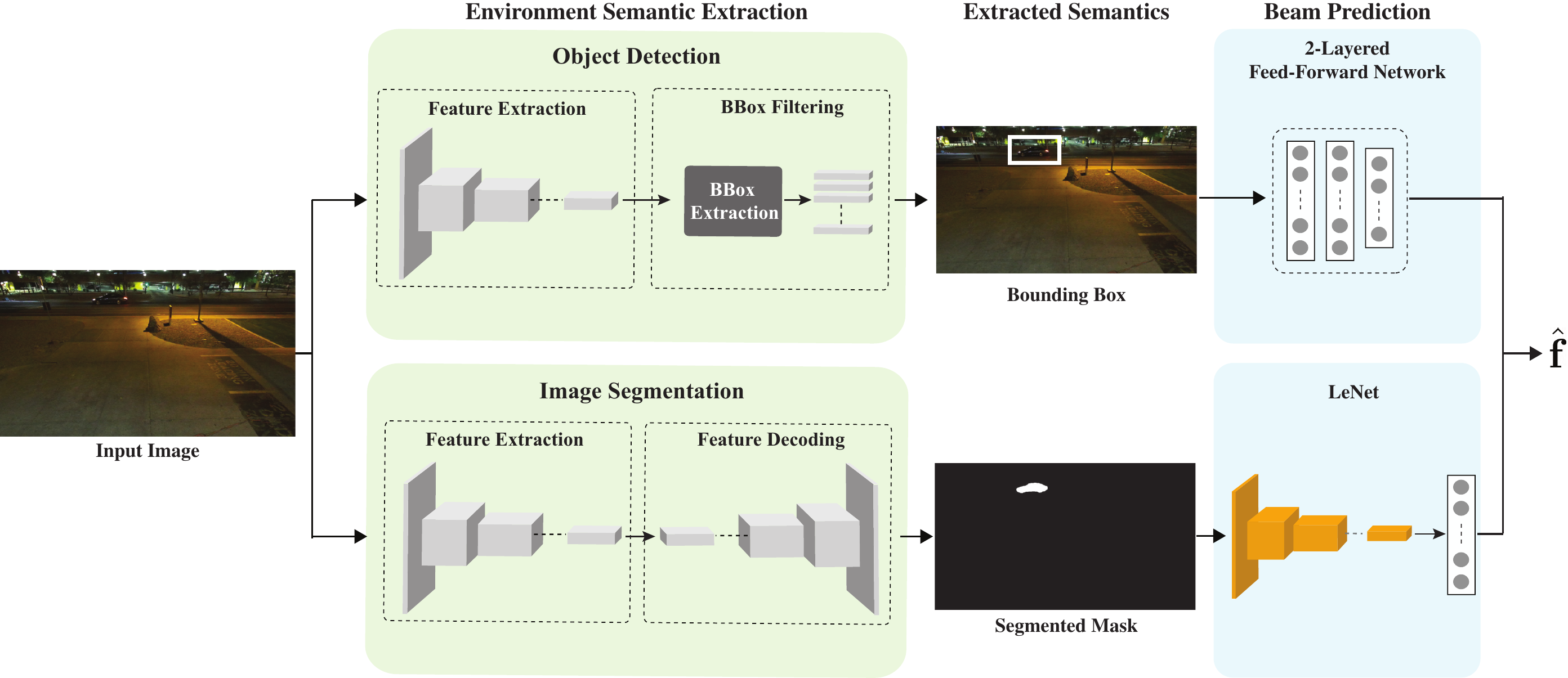}
	\caption{A block diagram showing the proposed solution for environment semantics-aided beam prediction task. As shown in the figure, the camera installed at the basestation captures real-time images of the wireless environment. We propose to first extract environment semantics (object masks, bounding boxes, etc.) from the RGB images. These extracted semantics can then be used to predict the optimal beam indices.    }
	\label{fig:beam_pred_soln}
\end{figure*}


\section{Environment Semantic Aided Beam Prediction: \\The  Key Idea and A Deep Learning Solution} \label{sec:prop_sol}
This section presents an in-depth overview of the proposed environment semantics-aided beam prediction solution. For this, we first present the key idea in Section~\ref{sec:key_idea} followed by the details of our proposed solution in Section~\ref{sec:ml_model}.

\subsection{The Key  Idea}\label{sec:key_idea}
Enabling highly-mobile mmWave/THz wireless communication applications requires overcoming some of the critical challenges associated with these high-frequency systems. One key challenge arises from the severe path loss associated with these high-frequency signals. In order to overcome this challenge, the mmWave/THz communication systems adopt large antenna arrays and use narrow directive beams in both the transmitter and receiver. However, selecting the optimal beams in these systems with large antenna arrays results in large beam training overhead, making it challenging to support high-mobility wireless communication applications. In general, the mechanism of directing the narrow beams can be viewed as focusing the wireless signal in a particular direction in space. The beam vectors theoretically divide the wireless environment (spatial dimension) into multiple (possibly overlapping) sectors. Therefore, if we have access to a pre-defined codebook, the beam prediction task can be re-defined as a classification task, where depending on where the user is located in the wireless environment, a particular beam index from the beam codebook can be assigned. 

In order to perform the beam classification task, it is imperative to understand and extract the exact user's location in the wireless environment. One promising way to achieve this is by utilizing additional sensing modalities such as vision, GPS, etc. The recent advancements in machine learning and computer vision, in particular, have provided capabilities to accurately identify and locate the objects of interest in the visual scene. For example, object detection models such as You Only Look Once version 7 (YOLOv7) can detect different objects in the visual data and provide bounding box coordinates. This work utilizes visual data to reduce the beam training overhead associated with mmWave/THz communication systems. Prior work on vision-aided beam prediction for mmWave/THz communication systems \cite{Charan_TxID, charan2021c } has primarily proposed solutions that provide the image as input directly to a convolutional neural network to predict the optimal beams. Although these initial approaches have provided promising results, the proposed solutions are computationally expensive, which further limits their adoption in the real world. In this work, we propose to first extract environment semantics from the images (captured at the basestation), such as image segmentation mask and bounding box information. The extracted semantics is then utilized to predict the optimal beamforming vector from a pre-defined beam codebook.

\subsection{Proposed Solution}\label{sec:ml_model}

In this subsection, we propose a two stage environment semantics aided beam prediction solution as shown in Fig.~\ref{fig:beam_pred_soln}. In the first stage, a machine learning model is deployed to extract the environment semantics in the form of either masks or bounding boxes of the objects of interest. The semantics are then utilized, in the second stage, to predict the beam index from a pre-defined beamforming codebook.

\textbf{Environment Semantics Extraction:} In the first stage, we adopt a  deep neural network (DNN) that takes the RGB image $\bX$ as input and provides the semantic representation for the RGB image, $\bS$, as the output. In general, the DNN must fulfill two essential conditions: (i) It should provide an accurate semantic representation and (ii) have a low computational footprint. In this paper, we adopt two state-of-the-art object detection models, YOLOv7 \cite{wang2022yolov7}, and MobileNet version 2  (MobileNetv2) \cite{sandler2018mobilenetv2} for generating these semantics. It is important to note that the YOLOv7 model has significantly more parameters than MobileNetv2. In addition, while YOLOv7 and MobileNetv2 have been designed specifically for object detection and generating bounding box coordinates of the detected objects, they can be adapted for image segmentation as demonstrated in \cite{mohamed2021insta}. To that end, we design a semantic segmentation head on top of a MobileNetv2 architecture for extracting the masks of the detected objects. Moreover, instead of training these models from scratch, we utilize pre-trained models as they can detect most of the relevant objects found in a wireless environment. To perform a detailed comparison between accuracy and latency, we generate two different types of semantics to represent the users in the environment, namely their binary masks $\bS{_\text{mask}}\in \mathbb{R}^{W \times H}$ and their bounding box vectors $\bS{_\text{bbox}}\in \mathbb{R}^{ 4 \times 1}$. The bounding box vector $\bS{_\text{bbox}} = [x_c, y_c, w,h]$ where $x_c, y_c, w,$ and $h$ are the $x$-centre, $y$-center, width, and height of the detected object, respectively. The extracted binary mask is then downsampled to the size $\hat{W} \times \hat{H}$ before it is provided as input to the DNN in the second stage. Unlike RGB images, the binary masks mostly retain their meaningful information, namely the positions and the shapes of the detected objects, when they are downsampled. Further, both the binary masks and bounding box vectors are normalized to the range $[0,1]$.

\begin{figure}[!t]
	\centering
	\includegraphics[width=0.95\linewidth]{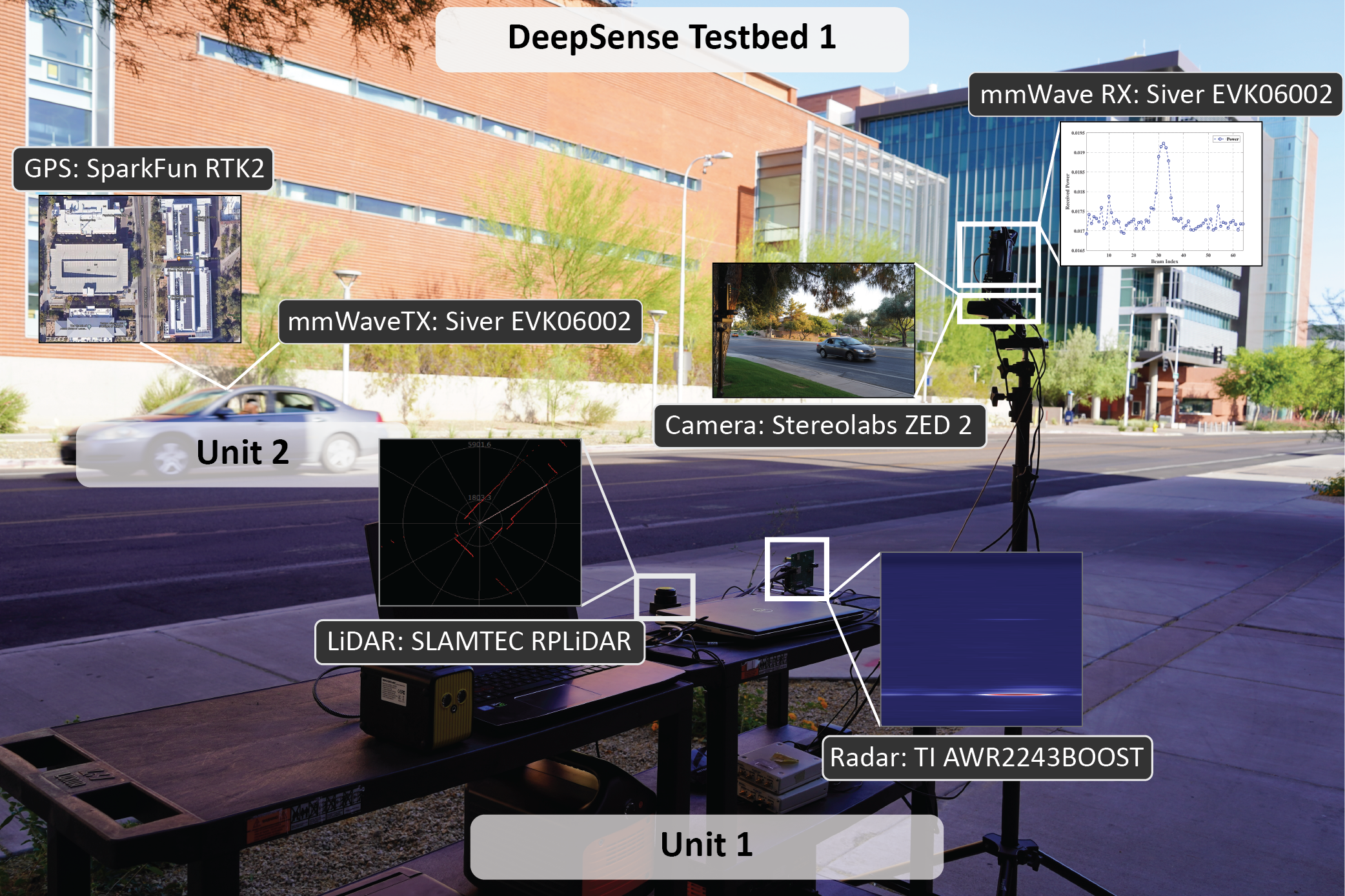}
	\caption{This figure presents the DeepSense 6G testbed 1 and the different sensing modalities. It consists of two units: Unit 1 (a stationary unit), acting as the basestation, and unit 2 (a vehicle), representing a mobile user.}
	\label{fig:deepsense_testbed}
\end{figure}

\textbf{Beam Prediction:} The second stage consists of a separate decision network that takes the semantic representation $\bS$ as the input and predicts the optimal beam index $ \hat{\mathbf f}$ as the output. Similar to the requirements of the semantics extraction network, the decision network is expected to (i) accurately predict the beam index and (ii) have a low computational footprint. We further observe that the structure of the semantic representation is significantly different for masks and bounding box vectors. Consequently, we adopt two different DNN architectures for the two modes of semantic representation utilized in this work.  

$1) \textit{ Fully Connected Neural Network for }\bS{_\text{bbox}}$: For the bounding box vectors, we use a $2$-layered fully connected neural network (FCNN) with $175$ neurons in each layer. FCNNs work well with structured data as they use the network's weights to model the relationship between every input element. Moreover, FCNNs make dense connections between the neurons of adjacent layers enabling them to learn more complex relationships between the input elements.

$2) \textit{ LeNet for }\bS{_\text{mask}}$: For images, however, convolutional neural networks (CNNs) have achieved better performance and robustness as they take advantage of the spatial correlation between the neighboring pixels. Therefore, for the binary masks, we adopt a simple CNN model (LeNet \cite{lecun1998gradient}) to predict the optimal beam index. The LeNet consists of 2 convolutional layers followed by two fully connected layers. The LeNet and the FCNN take masks and bounding vectors as inputs, respectively, and learns to predict the optimal beam indices. 


\begin{figure*}[!t]
	\centering
	\includegraphics[width=1.0\linewidth]{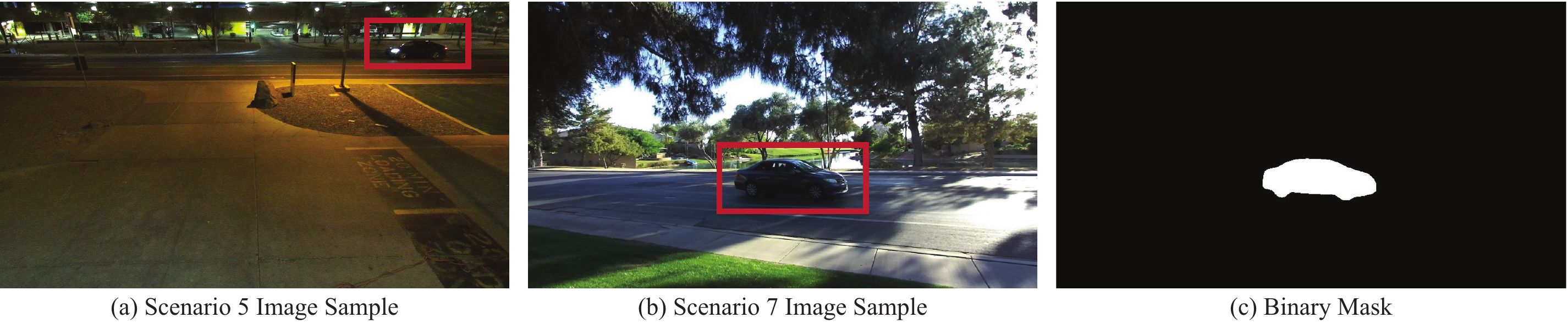}
	\caption{This figure presents the overview of the different data collection locations.  Fig. (a) and (b) present the visual data (RGB images) captured in scenario $5$ and $7$, respectively. We also show the corresponding bounding box of the mobile unit in the image.  Fig. (c) shows the corresponding mask of the mobile unit in the image. }
	\label{fig:deepsense_dataset}
\end{figure*}

\section{Testbed Description and Development Dataset}\label{sec:dataset}

To test the effectiveness of the proposed environment semantic-aided beam prediction solution, we utilize DeepSense 6G \cite{DeepSense} dataset. DeepSense 6G is a real-world multi-modal dataset developed for sensing-aided wireless communication applications. It consists of co-existing multi-modal data such as mmWave wireless communication, GPS data, vision, Radar, and LiDAR collected in a real-world wireless environment. In this section, we first present a brief overview of the DeepSense 6G testbed utilized during the data collection. Next, we present an analysis of the final development dataset used for developing and evaluating the proposed beam prediction solution. 

\textbf{DeepSense 6G: [Scenarios 5 and 7]} In this work, we adopt scenarios $5$ and $7$ of the DeepSense 6G dataset to evaluate the efficacy of the proposed solution. The hardware testbed and the example image samples from both scenarios are presented in Fig.~\ref{fig:deepsense_dataset}. The DeepSense testbed $1$ is utilized for this data collection, which consists of a stationary and mobile unit. The mobile unit (vehicle), acting as the transmitter, is equipped with a $60$GHz quasi-omni antenna and a GPS receiver to record the real-time location of the user. The stationary unit (basestation) is equipped with (i) an RGB camera and (ii) a $16$-element $60$GHz mmWave phased array. It uses an over-sampled predefined codebook of $64$ beams for receiving the transmitted signal. The data collected at each time instant consists of the GPS position of the user, RGB images, and the mmWave receive power vector. For further details, the reader is referred to \cite{DeepSense}, which describes the data collection testbed in detail.

\textbf{DeepSense 6G Development Dataset:}
We separately test the effectiveness of the proposed solution on scenarios $5$ and $7$ of the DeepSense 6G dataset. The measurements of scenario 5 and scenario 7 are collected at different times of the day in different locations as shown in Fig.~\ref{fig:deepsense_dataset}. In particular, scenario 7 measurements are collected in daylight whereas scenario 5 measurements are taken at night. In addition, both scenario 5 and scenario 7 contain many such images in which there are multiple mobile vehicle units but only one of them is transmitting to the basestation. Scenario 5 and scenario 7 contain 2300 and 854 samples respectively, which are further split into training, validation and testing samples with a ratio of 70/20/10 respectively.



\begin{table}[!t]
	\caption{Beam Prediction: Design and Training Hyper-parameters}
	\centering
	\setlength{\tabcolsep}{5pt}
	\renewcommand{\arraystretch}{1.2}
	\begin{tabular}{@{}l|cc@{}}
		\toprule
		\toprule
		\textbf{Parameters}                     & \textbf{Mask}  & \textbf{Bounding Box }            \\ \midrule \midrule
		\textbf{ML Model}                       & LeNet-5           & 2-layered MLP                   \\
		\textbf{Batch Size}                     & 64                  & 128                                 \\
		\textbf{Learning Rate}                  & $1 \times 10 ^{-3}$ & $1 \times 10 ^{-2}$ \\
		\textbf{Learning Rate Decay}            & epochs 10 and 20       & epochs 15 and 30  \\
		\textbf{LR Reduction Factor} & 0.1                 & 0.1                               \\
		\textbf{Total Training Epochs}          & 30                  & 50                              \\ \bottomrule \bottomrule
	\end{tabular}
	\label{tab_beam_pred_train_params}
\end{table}



\section{Performance Evaluation} \label{sec:perf_eval}
In this section, we study the performance of the proposed solution for the environment semantic-aided beam prediction task. For this, in Section~\ref{sec:exp_setup}, we present the details of the experimental setup adopted in this work. Next, we discuss the performance of the proposed solution in Section~\ref{sec:numerical_results}.

\subsection{Experimental Setup:} \label{sec:exp_setup}
In this work, we first extract the environment semantics which are then utilized to predict the optimal beam indices. As presented in Section~\ref{sec:prop_sol}, we utilize two different state-of-the-art object detection models (YOLOv7 and MobileNetv2) to extract the semantics from the visual data captured by the basestation. The two types of semantics (image mask and bounding boxes) extracted from the images require specialized machine learning-based models to perform the beam prediction task. For the binary mask-based approach, we design a CNN-based architecture (similar to that of LeNet). For the bounding box-based solution, we adopt a 2-layered fully-connected neural network. The proposed ML models are trained and validated on the task-specific dataset as presented in Section~\ref{sec:dataset}. The cross-entropy loss with the Adam optimizer is used to train the models. The details of the hyper-parameters used to fine-tune the models are presented in Table~\ref{tab_beam_pred_train_params}.

\textbf{Evaluation Metric:} The primary metric adopted to evaluate the proposed solution is the top-$k$ accuracy. Note that the top-k accuracy is defined as the percentage of the test samples where the optimal ground-truth beam is within the top-$k$ predicted beams. This work presents the top-1, top-2, and top-3 accuracies to evaluate the proposed solutions comprehensively.

\subsection{Numerical Results} \label{sec:numerical_results}
With the experimental setup described in Section~\ref{sec:exp_setup}, in this subsection, we study the beam prediction performance of the proposed solution.

\begin{figure}[t]
	\centering
	\subfigure[DeepSense Scenario 5]{\centering \includegraphics[width=1.0\linewidth]{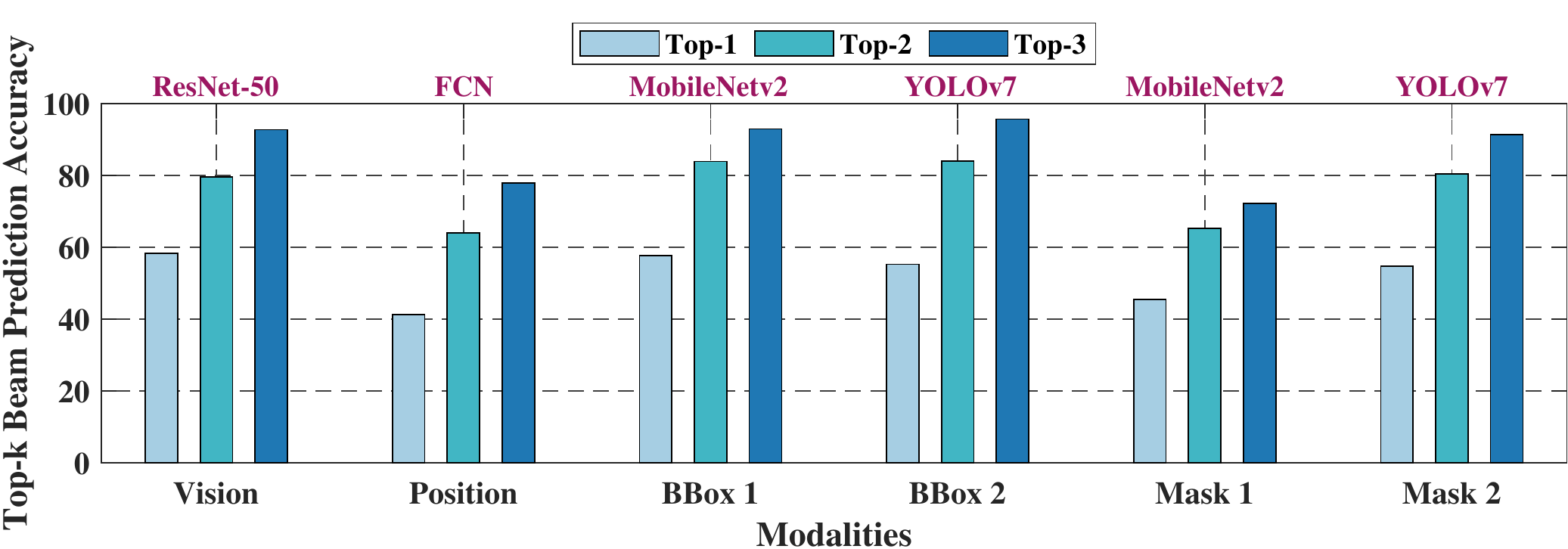}\label{fig:scenario5_beam_pred_acc}}
	\subfigure[DeepSense Scenario 7]{\centering \includegraphics[width=1.0\linewidth]{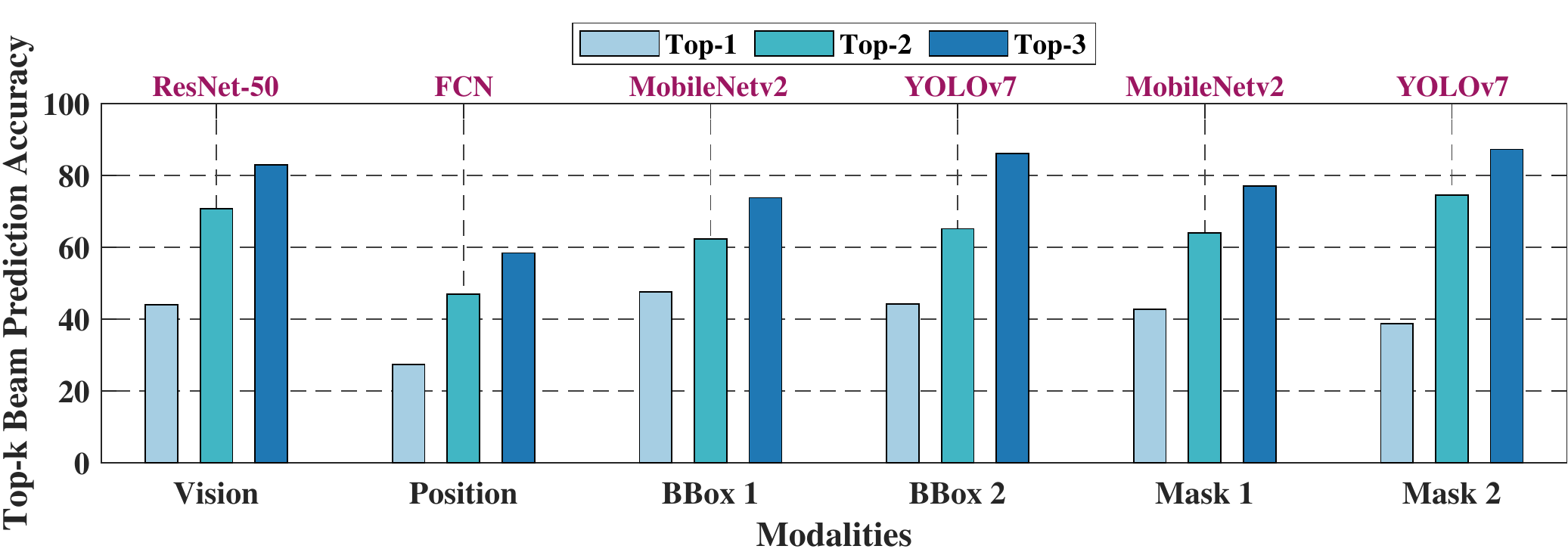}\label{fig:scenario7_beam_pred_acc}}
	\caption{This figure plots the top-k accuracies $(k \in (1,2,3))$ for the proposed environment semantics-aided beam prediction solution. We, further, plot the beam prediction accuracies for two prior work with visual and position data. It is observed the proposed bounding box-based solution (MobileNetv2) achieves similar or better performance than the prior vision-based solution.}
	\label{fig:acc_analysis}
\end{figure}

\textbf{Can environment semantics be utilized to predict the optimal beams?}
In order to perform a comparative study, as presented in Section~\ref{sec:prop_sol}, we utilize two state-of-the-art object detection models (YOLOv7 and MobileNetv2) to extract the environment semantics. To further facilitate a detailed study, we extract the image segmentation masks and the bounding boxes of the detected objects. In Fig.~\ref{fig:acc_analysis}, we present the top-$1$, top-$2$, and top-$3$ accuracies achieved by the different models utilized and the different extracted semantics for both scenarios $5$ and $7$ dataset. We also present the beam prediction accuracies of two prior work that utilize additional sensing data such as vision \cite{charan2021c} and position \cite{Morais22}. It is important to highlight here that different from our approach, the prior work with visual data provides the image directly as an input to a CNN. The beam prediction performance of these approaches will act as the baseline for our proposed solution. From Fig.~\ref{fig:acc_analysis}, it is observed that the bounding box-based solution can achieve similar beam prediction accuracy or even surpass the vision-based baseline solution. We observe similar performance trend for both the bounding boxes extracted from YOLOv7 and MobileNetv2. The proposed solution based on image segmentation mask achieves slightly lower performance as compared to the baseline vision and bounding box-based approach. This can be attributed to the fact that the generated masks were noisy in nature; a drawback that can be improved with further processing. However, both the mask and bounding box-based solutions were able to surpass the position-alone solution, highlighting the efficacy of utilizing environment semantics for predicting the optimal beam indices in mmWave/THz communication systems. 

\begin{figure}[!t]
	\centering
	\includegraphics[width=0.95\linewidth]{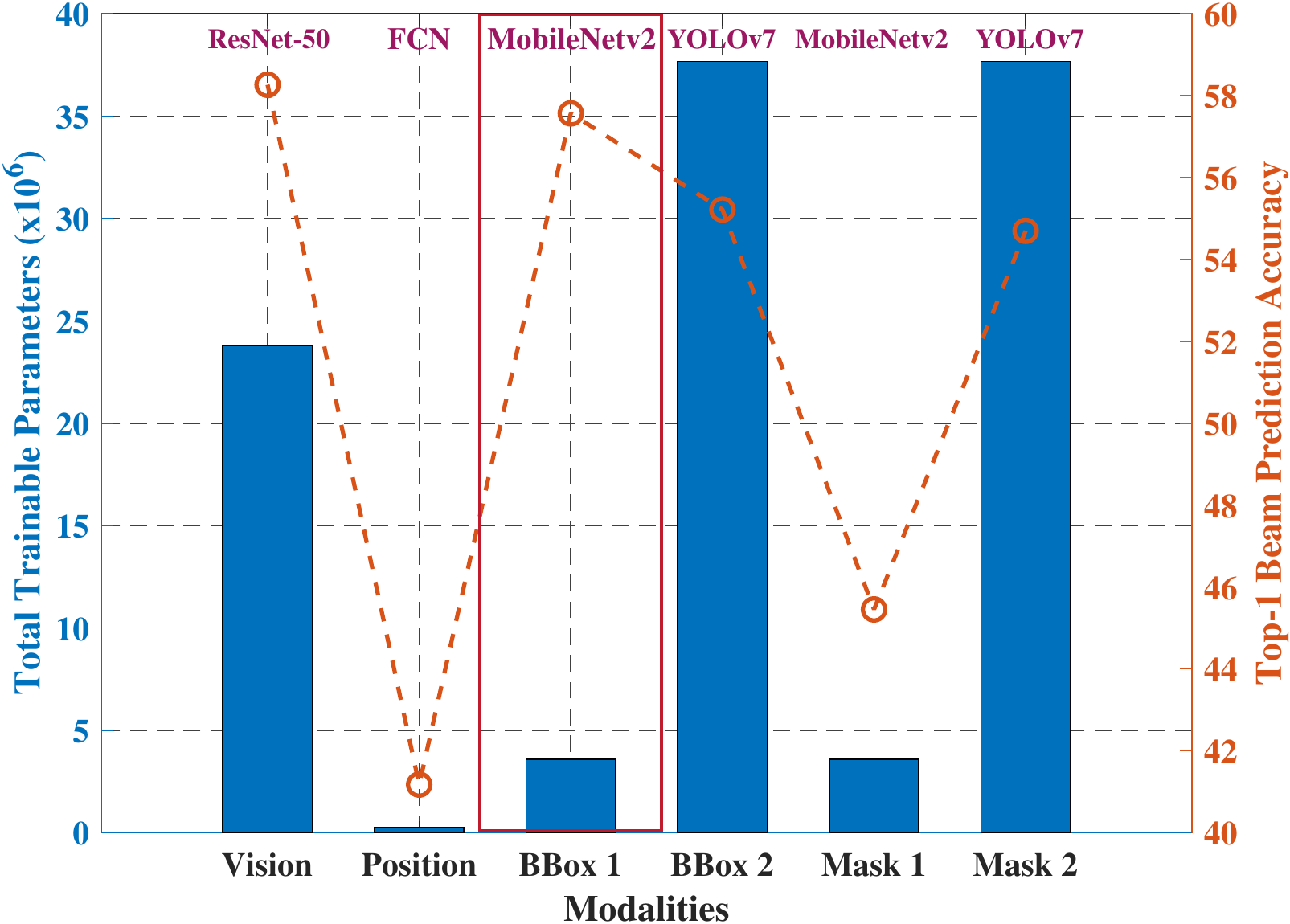}
	\caption{This figure plots the total number of trainable parameters versus the top-$1$ beam prediction performance for the different approaches. It shows that the bounding box-based solution can achieve similar top-1 beam prediction accuracy while using less than $1/6^{th}$ of the parameters when compared with the baseline vision-based approach.  }
	\label{fig:computational_cost}
\end{figure}
\textbf{Trade-off between computational cost and model performance:} The final adoption of any proposed solution depends on achieving both high accuracy and low latency. As shown in Fig.~\ref{fig:acc_analysis}, the proposed environment semantics-aided beam prediction (bounding box-based) solution achieves high prediction accuracy compared to the baseline vision-based solution. In Fig.~\ref{fig:computational_cost}, we plot the computational requirement versus the top-$1$ beam prediction performance for the different approaches. Although the position-based beam prediction approach has the lowest footprint ( in terms of model parameters), it achieves the lowest beam prediction performance. This can be attributed to the inherent errors in GPS data, and access to accurate positions in the future might help improve the performance. It is also interesting to note that the bounding box-based approach (extracted using MobileNetv2) achieves similar performance as the baseline vision-based solution with only a fraction of the trainable model parameters. These results highlight the computational efficacy of the proposed environment semantics-based beam prediction solution. It also emphasizes that there is an inherent trade-off between inference accuracy and achievable latency, which must be considered during network design.

\section{Conclusion}\label{sec:conc}

This paper develops a two stage deep learning based solution for fast and accurate mmWave/THz beam prediction. In the first stage, the semantics are extracted from the visual data. These semantics are then used for beam prediction in the second stage. We evaluate the proposed solution on the DeepSense 6G dataset. Using bounding box vectors from MobileNetv2, the proposed solution achieves similar top-1 beam prediction accuracy while using less than $1/6$ of the parameters when compared with the vision based approach, highlighting a promsing solution for real-world systems.

\bibliographystyle{IEEEtran}

\end{document}